\documentclass[prl,aps,showpacs,twocolumn]{revtex4}
\usepackage{graphicx}
\usepackage{amsmath}
\usepackage{amsfonts}
\usepackage{amssymb}
\usepackage{verbatim}
\usepackage{color}
\usepackage{ulem}
\providecommand{\U}[1]{\protect \rule{.1in}{.1in}}

\begin{document}
\title{The Helical Superstructure of Intermediate Filaments}
\author{Lila Bouzar$^{a}$, Martin Michael M\"{u}ller$^{b,c}$, Ren\'{e} Messina$^{b}$, 
Bernd N\"{o}ding$^{d}$, Sarah K\"{o}ster$^{d}$, Herv\'{e} Mohrbach$^{b,c}%
$, Igor M. Kuli\'{c}$^{c}$}
\affiliation{$a$\ Laboratoire de Physique des Mat\'eriaux, USTHB, BP~32 El-Alia Bab-Ezzouar, 16111 Alger, Algeria}
\affiliation{$b$\ Laboratoire de Physique et Chimie Th\'eoriques - UMR 7019, Universit\'e de Lorraine, 1
boulevard Arago, 57070 Metz, France}
\affiliation{$c$\ Institut Charles Sadron, CNRS-UdS, 23 rue du Loess, BP 84047, 67034
Strasbourg cedex 2, France}
\affiliation{$d$\ Institute for X-Ray Physics, University of Goettingen, Friedrich-Hund-Platz 1, 37077 G\"{o}ttingen, Germany}

\begin{abstract}
Intermediate filaments are the least explored among the large cytoskeletal elements. 
We show here that they display conformational anomalies in narrow microfluidic channels.
Their unusual behavior can be understood as the consequence of a
previously undetected, large scale helically curved superstructure. 
Confinement in a channel orders the otherwise soft, strongly fluctuating helical
filaments and enhances their structural correlations, giving rise to
experimentally detectable, strongly oscillating tangent correlation functions. We
propose an explanation for the detected intrinsic curving phenomenon - an elastic shape instability that we call autocoiling. The mechanism
involves self-induced filament buckling via a surface stress located at the outside of the
cross-section. The results agree with ultrastructural findings and rationalize for the commonly observed looped intermediate filament shapes.
\end{abstract}

\maketitle


The integrity and dynamics of biological cells delicately depend on the
mechanical response of their cytoskeleteton consisting of actin filaments, 
microtubules and intermediate filaments (IFs) \cite{Alberts,Amos}. While they
have been experimentally probed in many different ways, a full
understanding of their properties still remains a challenge. 
This is particularly true for the IFs - the least studied of. Experiments probing their elastic response have shown that they are the most flexible and extensible of the filaments of the cytoskeleton \cite{Muecke2004,Kreplak2005,Kreplak2008}. Bio-filaments are usually modeled as semiflexible polymers characterized by their persistence length.
For vimentin, a prominent member of the IF family, this model was
used in combination with various experimental methods to
determine values for the persistence length of a few $\mu$m  \cite{Winheim2011,Schopferer2009,Muecke2004}.

\begin{figure}[hptb]
\includegraphics[width=0.5\textwidth]{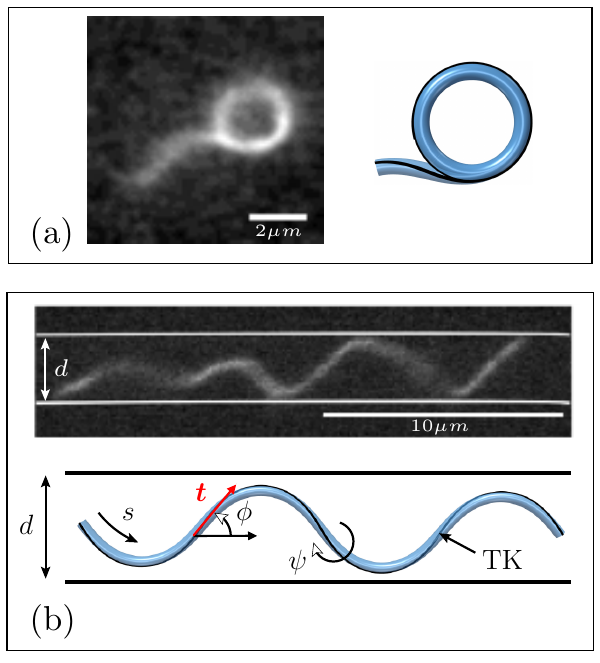}
\caption{
(a) 
Left panel: Vimentin filament confined in a slit occasionally coiled up in a transient ring. 
Right panel: Model three dimensional helical filament getting confined to a plane forming a ring 
as an energy minimum, see Eq. \eqref{eq:E} and text.   
(b)
Upper panel: Typical oscillatory shape of an intermediate filament in a quasi two-dimensional microfluidic channel
characterized by a lateral confinement $d$ 
\cite{KosterFI}.
Lower panel: Geometrical setup of the model filament. The tangent vector $\boldsymbol{t}$ evolves with arc-length $s$ along the filament. Its direction is given by the angle $\phi(s)$, measured from the horizontal. $\psi(s)$ denotes the twist angle represented by a black ribbon on the filament's surface. The variation of the twist at curvature inversion points gives rise to conformational defects called twist kinks (TK).
}
\label{fig:intro_manip}
\end{figure}

In most of the experiments the IFs interact strongly with a substrate 
(AFM and EM) and the extracted physical
properties depend on the substrate properties the filaments adhere
to. A closer look at the experimental micrographs \cite{Winheim2011,Schopferer2009,Muecke2004} 
reveals shapes that resemble sinusoidal waves, loops or
circular arcs  reminiscent of 
helices confined to a 2D substrate
\cite{Gimoon,Lila}. This leads to the suspicion that IFs are not  simple enough to be described by a semiflexible chain. To eliminate possible artifacts from
adsorption, we have studied individual IFs in
quasi two-dimensional microfluidic channels where the filaments, despite
geometric confinement, are free to rearrange.
Averaging over a number of different filaments a
persistence length close to 2$\,\mu$m was found previously \cite{KosterFI}. However, a deeper inspection
of the \textit{individual} filament data, that we will present in
this paper, reveals an anomalous behavior which is incongruent
with the expected behavior of a semiflexible polymer. For instance, we observe the transient formation of rings and oscillatory shapes (see Fig. \ref{fig:intro_manip}). However, the most prominent manifestation of this anomaly manifests itself in a strongly
oscillating tangent correlation function for individual filaments,
in sharp contrast to the behavior characteristic for a semiflexible chain under lateral
confinement \cite{Odijk}. In this paper we will show that the data can be completely
rationalized by assuming that IFs behave like squeezed helical
filaments under lateral confinement.

A central observable in our study is the tangent correlation function, $G(s)$, that 
provides crucial conformational information about the filament's microstructure.
More specifically, it is defined as  

\begin{eqnarray}
\label{eq:G_def}
G(s)= \left \langle \overline { \cos \phi(s) } \right \rangle, 
\end{eqnarray}
where $\phi(s)$ is the tangent angle of the filament at the arc length position $s$, 
see sketch in Fig. \ref{fig:intro_manip}(b). 
$\left \langle \dots \right \rangle$ in Eq. \ref{eq:G_def} denotes thermal average and $\overline{(\dots)}$ the spatial average along
the contour length ($L$), i.e. $\overline{f[\phi(s)]}=\frac{1}{L-s}\int_{0}^{L-s}du f[\phi(u+s)-\phi(u)]$, 
see also Fig. \ref{fig:intro_manip}(b).

\begin{figure}[ptb]
\includegraphics[width=8.5cm]{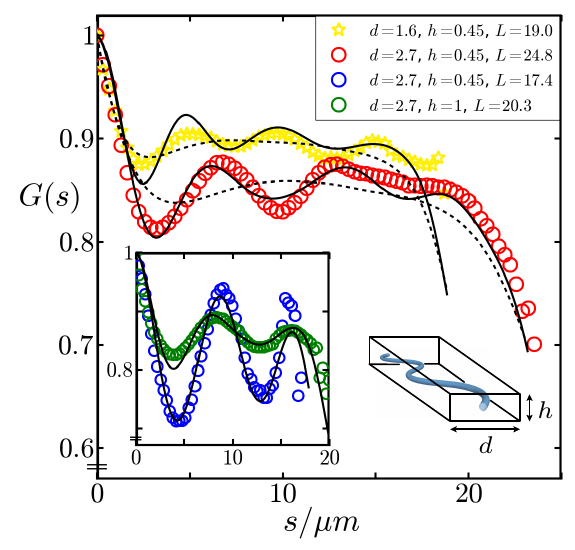}
\caption{
Tangent correlation functions $G(s)$ of vimentin filaments (experimental data represented by symbols) confined in a quasi two-dimensional microfluidic 
channel of width $d$ and height $h$ given in microns (as well as $L$), see bottom right inset.
Bottom left inset: (i) Data for $h=1 ~\mu \rm m$ (green circles). (ii) Sample exhibiting strong and persisting oscillations in
$G(s)$ (blue circles). 
Solid (dashed) lines stem from MC simulation data of filaments in two dimensions (i.e., $h=0$) with (without) twist. 
}
\label{fig:G_of_s}
\end{figure}

Revisited (unpublished) experimental data on vimentin filaments whose set up is described in detail elsewhere \cite{KosterFI} 
are displayed in Fig. \ref{fig:G_of_s}. Quasi two-dimensional microfluidic channels of height $h$ and varying 
lateral confinement $d$ were employed to confine these filaments.
Upon limiting the thermal average of $\overline {\cos \phi(s) }$  to an \textit{individual} filament,
striking oscillations in $G(s)$ set in, see Fig. \ref{fig:G_of_s}.  
At prescribed height ($h=0.45 ~\mu \rm m$), the two profiles 
in the main body of figure \ref{fig:G_of_s} (red and yellow symbols) correspond to two different lateral confinements 
($d=1.6~\mu \rm m$ and $d=2.7~\mu \rm m$). 
$G(s)$ increases with decreasing $d$ (i. e., with increasing degree of lateral confinement) as expected, 
whereas the oscillation amplitudes as well as their associated wave length decrease.   
These features remain robust upon changing the channel height to $1~\mu \rm m$, see 
green circles inset of Fig. \ref{fig:G_of_s}. 
It is also possible to observe even stronger and more persistent oscillations, although such conformations are 
rather rare events, see data with blue circles in inset of Fig. \ref{fig:G_of_s}. 

In the following, we are going to rationalize these experimental observations by means of Monte Carlo (MC)
simulations as well as analytical theory. The underlying idea is that confined IFs can be modeled as 
a  helical superstructure trapped on a flat surface and subject to an additional lateral confinement. 

Consider a semi-flexible helical filament of length $L$ becoming confined to a plane. 
Its elastic energy is given by \cite{Gimoon,Lila}
\begin{eqnarray}
\label{eq:E}
E & = &\frac{1}{2}\displaystyle \int_{0}^{L}[B\left(  \phi^{\prime}-\omega_{1}\sin
\psi \right)  ^{2}+C\left(  \psi^{\prime}-\omega_{3}\right)  ^{2}
\nonumber \\
&&  +B\omega_{1}^{2}\cos^{2}{\psi}]ds\;, 
\end{eqnarray}
where $\phi^{\prime}(s)=:\kappa(s)$ stands for the curvature, $\psi$ designates the twist angle and 
$\psi^{\prime}(s)$ the twist with $s$ being the arc-length, see also Fig. \ref{fig:intro_manip}(b).
The constants $B$ and $C$ in Eq. \eqref{eq:E} are the bending and torsional stiffness,
respectively, $\omega_{1}$ and $\omega_{3}$ are the preferred
curvature and twist of the unconfined three-dimensional helical filament \cite{helices}.

The filament ground state stemming from Eq. \eqref{eq:E} obeys two coupled equations:
The pendulum-like equation 
(i) $\psi^{\prime \prime}+\frac{B\omega_{1}^{2}}{2C}\sin(2\psi)=0$
and (ii) $\kappa=\omega_{1}\sin \psi$ indicating that curvature is slaved by the twist angle
in contrast to the unconfined three-dimensional case (where both decouple).
In general, depending on the material parameters, a rich variety of equilibrium shapes resembling
loops, waves, spirals or circles exist \cite{Lila}. 
As a matter of fact, these shapes can be seen as the result
of interacting repulsive conformational defects corresponding 
to curvature inversion points. In terms of twist, such defects originate from 
a rapid variation of $\psi(s)$ (reminiscent of a kink) and are called twist-kinks (TKs).
A relevant dimensionless parameter is $\gamma=\frac{4\omega
_{1}^{2}B}{\pi^{2}\omega_{3}^{2}C}$ which measures the ratio of bending and twisting energy. 
For $\gamma>1$, the ground state approaches a twist-kink-free circular arc of radius
$1/\omega_{1}$, see Fig. \ref{fig:intro_manip}(a). 
For $\gamma<1,$ the filament can be populated
by twist-kinks whose density is limited by their repulsion 
\footnote{The number of twist-kinks can fluctuate at a finite
temperature. The translational zero mode of the injected and
moving twist-kinks leads to a chain with anomalously large
"hyperflexibility" \cite{Gimoon}.A remarkable analogy can be established with \textit{dislocations} in solids,
where (plastic) deformation is facilitated by the production of such energetically cheap 
linear defects in crystals \cite{Friedel}. }.
\begin{figure}[ptb]
\centering
\includegraphics[width=8.5cm]{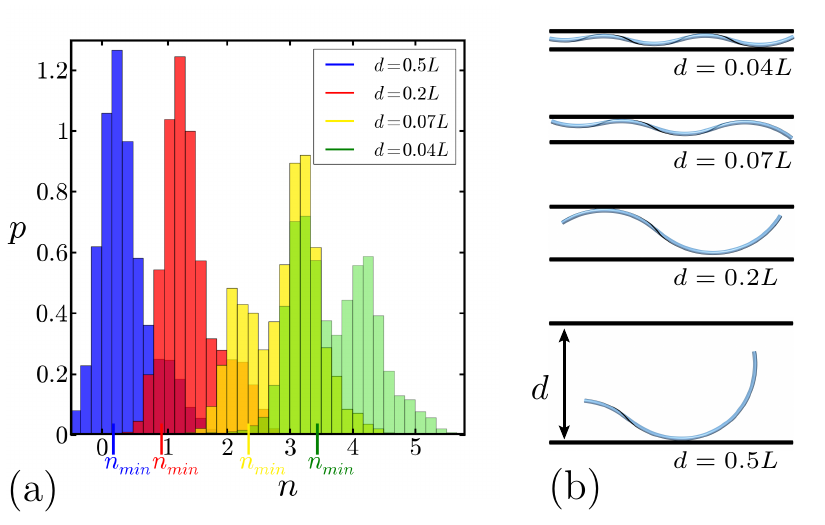}
\caption{(a) Simulation data for the probability density of the number of twist-kinks $n$ for different values of 
the lateral confinement $d$ with $\omega_{1}=\pi$, $\gamma=5$, $B=10$,  $C=0.16$, 
(with $k_BT=L=1$). 
(b) Low tempertaure snapshots  of a filament close to its ground state for the same
values of $d$ as in the histogram illustrating the twist-kink injection.}
\label{fig:TK_MC}
\end{figure}
Extensive MC simulations based on the Hamiltonian of Eq. \eqref{eq:E} 
have been carried out in order to explore the conformations of confined filaments 
in full detail.  
The filament is modelled as a discrete chain made up of $N$ monomers (i.e., segments) of fixed length $b$
\footnote{
Each monomer is characterized by four variables $\left(  x_{i},y_{i},\phi
_{i},\psi_{i}\right)  $ with $i=1\dots N$. The positions $x_{i}=x_{i-1}%
+\cos \phi_{i-1}$ and $y_{i}=y_{i-1}+\sin \phi_{i-1}$ stand for the Cartesian
coordinates in the plane. One Monte Carlo step consists of $N$ attempted local rotations and torsions of each monomer
sequentially and a single attempt of a global translation (only relevant in
the presence of the confining channel), rotation \cite{Madras} and torsion.
Standard Metropolis scheme is then used to generate a canonical ensemble
\cite{Allen}. Typically, $10^{6}$ Monte Carlo steps were employed for a
simulation run and about the second half of the steps was used to gather
statistics.}. 
Our best matching MC data for $G(s)$ can be found in Fig. \ref{fig:G_of_s}.
Interestingly, taking into account the twist via Eq. \eqref{eq:E} corroborates
the experimentally observed oscillations by employing consistent material parameters 
\footnote{
(i) For simulation data with finite twist, the elastic constants are $B/(k_B T) \simeq 50~\mu$m and
$C/(k_B T) \simeq 3~\mu$m except for the corresponding experimental data with blue circles  
where $C/(k_B T) \simeq 10~\mu$m.  The geometrical parameters are typically 
$\omega_{1}^{-1}\simeq  2~\mu$m of $\omega_{3}^{-1} \simeq 1 ~ \mu$m 
(ii) For twist-free filaments (i.e., $\omega_{1}=\omega_{3}=C=0$),  
$B/(k_B T) \simeq 5 ~ \mu$m. 
}, see Fig. \ref{fig:G_of_s}. 
These results suggest that the unconfined vimentin possesses a helical 
superstructure characterized by a radius $\sim 0.4~ \mu$m and 
pitch $\sim 4~ \mu$m and a bending persistence length $B/(k_B T) 
\sim 50 ~\mu \rm m$ \footnote{Note that the value of $B$ is much higher than determined previously from multi-filament averaging - a procedure that neglects single filament correlations and measures an ``apparent'' persistence length.}.
The microscopic origin of this helix is still unclear but a plausible 
mechanism is proposed at the end of this paper. In contrast, for twist-free semiflexible chains,
no oscillations emerge in $G(s)$ where the usual Odijk behavior \cite{Odijk} is recovered, see dashed lines in Fig.~\ref{fig:G_of_s}. 

To deepen our understanding of the role of the internal twist, that is a hidden degree of freedom in the experiments, 
we further utilize the simulation data to analyze the filament conformation for various degrees of lateral confinement.  
A useful quantity is the probability density of twist-kinks occurrence, $p(n)$,   
where $n$ is the number of  twist-kinks which is adequately defined as $n= \frac{1}{\pi} \int_0^L  \psi'(s) ds$. 
Probability density profiles $p(n)$ for varying lateral confinement  $d$ are depicted in Fig. \ref{fig:TK_MC}(a).
On average, the number of twist-kinks increases with decreasing $d$, see Fig. \ref{fig:TK_MC}(a). 
Computing the corresponding $G(s)$ (not shown) clearly indicates that the average 
number of twist-kinks $n$ also describes the number of extrema encountered in $G(s)$. 
Thermal fluctuations induce injections and ejections of twist-kinks at the ends of the filament,
which qualitatively explains the fading of the oscillations in $G(s)$ shown in Fig. \ref{fig:G_of_s}. 
Concomitantly, a finite temperature induces a broadening in $p(n)$.     

\begin{figure}[ptb]
\centering
\includegraphics[width=0.48\textwidth]{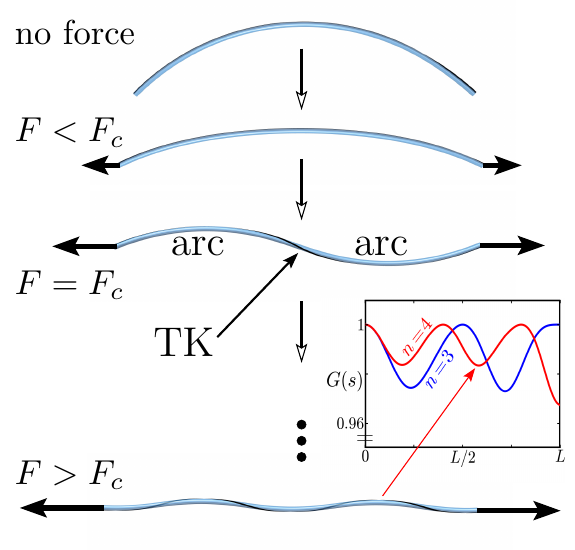}\caption{
Evolution of a ground state of an elastic filament under an external pulling force.
Inset: Corresponding tangent correlation functions for $n=3$ (blue) and
$n=4$ (red) twist-kinks (TKs). 
}
\label{fig:force}
\end{figure}

To shed more light on the underlying physical mechanisms of confinement-mediated 
twist-kink production let us consider a more tractable analytical model. It consists in replacing the confinement by a pulling force at zero temperature, see Fig.\ref{fig:force}. Although this systems is different it illustrates the twist-kink injection mechanism in a phenomenological manner. 
In the presence of the external force $F$ and for small deformations ($\varphi \ll1)$ the elastic energy becomes
\begin{eqnarray}
E_F \simeq  E + \frac {F}{2} \int_0^L \phi^{2}(s)ds.
\end{eqnarray}
Consider the regime $\gamma\gg1$  which has as a ground state (in absence of a force) a circular arc $\phi(s)=-\omega_{1}s$
and energy $E_0=CL\omega_3^2/2$, see Fig. \ref{fig:force}.
The formation of a twist-kink is energetically favorable when  
the pulling force exceeds a critical value $F_{c}=\frac{64}
{L^{3}\omega_{1}}\sqrt{BC}$. The latter is the result of comparison of the 
elastic energy of a twist-kink-free filament ($E_0+\frac{\omega_{1}^{2}L^{3}}{24}F$)  
with that of a filament containing a single twist-kink located at the midpoint
($E_0+\omega_{1}\sqrt{BC}
+\frac{\omega_{1}^{2}L^{3}}{96}F$), see Fig. \ref{fig:force}.  
In the opposite, small force regime ($F<F_c$), the filament deforms like a stretched elastic arc,
see Fig. \ref{fig:force}.

Above $F_c$, one or more twist-kinks are formed, see  Fig. \ref{fig:force}. The required force to nucleate $n$ twist-kinks is given 
by $F_{c}(n)=\frac{48\sqrt{BC}}{L^{3}\omega_{1}}\frac{n^{2}\left(
n+1\right) ^{2}}{2n+1}$. 
Note, that the kinks mutually repel each other giving rise to an ordered
one-dimensional crystal-like structure of $n$ twist-kinks separated by circular arcs of
switching curvature, see Fig. \ref{fig:force}.
Expanding the chain conformation in Fourier modes, and keeping only the dominant one we extract the tangent correlation function:  
\begin{eqnarray}
\label{eq:G_force}
G(s)\approx1-\frac{16\ell^{2}\omega_{1}^{2}}{\pi^{4}}\sin^{2}\left(  \frac{\pi
s}{2\ell}\right)  \left(  1+\frac{\ell \sin \left(  \pi s/\ell \right)  }
{\pi \left(  L-s\right)  }\right)
\end{eqnarray}
with $\ell=L/(n+1)$ representing the distance between two adjacent twist-kinks. 
Typical profiles of $G(s)$, displaying oscillations similar to the confinement case, are shown in Fig. \ref{fig:force}. 
Simulation results with external force, not shown here, corroborate these findings.   

Thus experimental and theoretical results strongly suggest that vimentin 
filaments have a helical superstructure, indicated by remarkable oscillations 
in the tangent correlation function in confined geometry. 

\begin{figure}[ptb]
\centering
\includegraphics[width=0.45\textwidth]{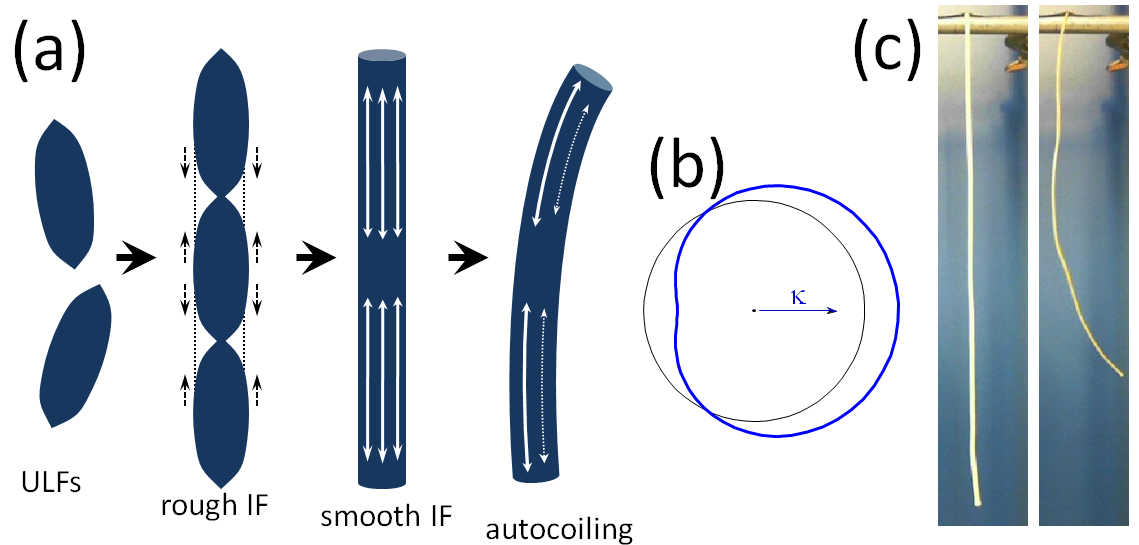}\caption{a) Stages of
assembly of IFs and their autocoiling instability. b) The cross-section
deformation of the bent circular rod. c) The autocoiling of drying spaghetti.
}
\label{fig:autocoiling}
\end{figure}

The molecular origin of the helical shape of vimentin filaments is unclear at this
point. Oscillations in the tangent correlation functions of individual actin filaments were also previously observed \cite{KosterActine1,KosterActine2} and helical superstructures of
microtubule reported \cite{MT1,MT2}. Thus, the question arises which general physical mechanism could be at the origin of the curved/helical shapes of those biofilaments.
Here we propose a mechanism  based on self-buckling, for which we suggest the name \textit{autocoiling}, see Fig.~\ref{fig:autocoiling}. It is a generic feature of filaments exposed to surface stresses that induce a broken symmetry and thus curved states. It  can be witnessed in the mundane example of a drying spaghetti, see Fig.~\ref{fig:autocoiling}(c). The outer layer of the spaghetti dries and shrinks faster and induces a buckling
stress on the (transiently) more swollen core. 

For vimentim, there are a few hints towards a surface stress from its molecular structure.
During the assembly of its constituent monomers the monomers form highly elongated coiled-coil dimers, then tetramers and at
an intermediate stage give rise to an approximately ``spindle shaped''
32-mer (see images in Ref.~\cite{Herrmann1999})---the so called unit length filament (ULF)
\cite{helical_intermediatefilaments}. These ULFs then assemble
into long filaments. The initially rugged, "rough" filaments of
spindle-like ULF subunits, longitudinally anneal and the filaments undergo a maturation phase during which the surface smoothens Ref.~\cite{Herrmann1999,helical_intermediatefilaments}. We suggest that the spindle shape and the finite lateral thickness of
the ULF can be understood as originating from
double-twist-frustration caused by chiral (cholesteric)
interactions of the coiled-coil alpha helices aligned along the
axis. As proposed by Grason and Bruinsma \cite{Bruinsma}, a bundle
of chiral objects displays a finite size as the chains on the outside are progressively more
tilted than those on the inside. This tilt causes a surface stress and
axial shortening of the outer chains with respect to the inner ones. 
As the simplest possible model for self-buckling under a surface stress we consider an isotropic incompressible elastic rod of radius $R$ and length $L$. Its energy is $U=\frac{1}{2}\int dV\sigma_{ij}\varepsilon_{ij}+\lambda
\Delta S$ where $\sigma_{ij}$ and $\varepsilon_{ij}$ are the bulk stress and strain tensor respectively. The constant $\lambda>0$ is an isotropic tensile surface stress (surface energy density).  
$\Delta S$ denotes the variation of the surface after deformation. When negative, the term $\lambda\Delta S$ can compete with the positive elastic bulk energy. 

Under a pure bending deformation with a constant curvature  $\kappa$ of the rod, the cross-section is slightly deformed (see \ref{fig:autocoiling}(a)) and $\Delta S<0$ as deduced from the deformation field of a bent rod \cite{Continuum}. It can be shown that $\frac{\Delta S}{S_{0}}\approx-\frac{3}{16}R^{2}\kappa^{2}$ where $S_{0}=2\pi LR$ is the surface of the undeformed rod. The total surface+elastic energy can then be written as 
$U=\frac{1}{2}B_{e}(\lambda)L\kappa^{2}$ with an effective bending modulus that has contributions from both the bulk and the surface stress of the rod : $B_{e}(\lambda)=\frac{\pi}{4}YR^{4}-\lambda\frac{3\pi}{4} R^{3}$ where $Y$ is the Young modulus. For sufficently large $\lambda$ the stiffness $B_e(\lambda)$ vanishes and the rod becomes unstable. 
This is the signature of a spontaneous broken symmetry of the rod, i.e., self-buckling. 

Beyond the curvature, the origin of helical torsion of vimentin,
its magnitude and handedness deserve further theoretical and
experimental investigations. At this point we speculate that the
handedness and the pitch of IFs are inherited from the  chiral inter-alpha-helix interaction on the monomer level.

In summary, we report experimental and theoretical evidence for a helical
superstructure in intermediate filaments. Lateral confinement in a quasi-2D channel 
orders the otherwise soft and strongly fluctuating twist-kink defects and leads to experimentally detectable, oscillating tangent correlation functions. The underlying physical mechanism is the injection of low energy defects, that can be seen as a new deformation mode for confined helical filaments. This behavior is reminiscent of the plastic deformations mediated by dislocations in solids \cite{Friedel}. 
The self-buckling mechanism suggested to be at the origin of the helical states of vimentin could also be common in other biological filaments beyond IFs. More generally, tensile surface stress could be induced by 
any mismatch between the surface and the bulk of the filament, like e.g. the surface tension between the ordered water and ions at the outer layers of the filament that display mismatched osmotic and Maxwell stresses between the interior and exterior. 

The biological meaning of an intrinsically stiffer but coiled helical structure is apparent. A coiled structure resists small forces by uncoiling and displays high compliance without damage up to the
point of stronger elongation. This reinforces the IFs' natural
role as supporting mechanical elements and protective stress absorbers of the cell.

\end{document}